\documentclass[11pt,a4paper]{article}

\usepackage[utf8]{inputenc}
\usepackage[T1]{fontenc}
\usepackage{lmodern}
\usepackage[english]{babel}
\usepackage{amsmath,amssymb,amsfonts}
\usepackage{graphicx}
\usepackage{xcolor}
\usepackage{booktabs}
\usepackage{tabularx}
\usepackage{multirow}
\usepackage{hyperref}
\usepackage{natbib}
\usepackage{lineno}
\usepackage{microtype}
\usepackage{geometry}
\usepackage{caption}
\usepackage{subcaption}
\usepackage{enumitem}
\usepackage{float}
\usepackage{pdflscape}
\usepackage{authblk}

\hypersetup{
    colorlinks=true,
    linkcolor=blue!60!black,
    citecolor=blue!60!black,
    urlcolor=blue!60!black,
    pdftitle={Generative AI impacts on intra-urban inequality and skill premium in Beijing},
}

\modulolinenumbers[5]

\title{Generative AI impacts on intra-urban inequality and skill premium in Beijing}

\author[1,\dag]{Xiliu He}
\author[1,\dag]{Haoxiang Zhao}
\author[1]{Mingyi Ma}
\author[1]{Edward Wenchuan Lai}
\author[1]{Koei Enomoto}
\author[1]{Anni Hu}
\author[1]{Jiatong Li}
\author[2]{Lingyun Chu}
\author[1,3,\ddag]{Yuan Lai}

\affil[1]{School of Architecture, Tsinghua University, Beijing, China}
\affil[2]{ZODA LAB, Hangzhou, China}
\affil[3]{Technology Innovation Center for Smart Human Settlements and Spatial Planning \& Governance,
Ministry of Natural Resources, Tsinghua University, Beijing, China}

\date{\today}

\begin{document}

\maketitle

\noindent\textsuperscript{\dag}These authors contributed equally to this work.\\
\textsuperscript{\ddag}Corresponding author. E-mail: yuanlai@tsinghua.edu.cn

% ─── Abstract ───
\begin{abstract}
Generative artificial intelligence (GenAI) is the first automation wave to reach high-cognitive tasks at scale, yet its effects on intra-urban inequality remain largely unknown. Using 5 million job postings from Beijing (2018--2024), we construct a neighborhood-level GenAI Exposure Index by aggregating task-level assessments from five leading large language models. We examine the spatial, structural and causal mechanisms of this shock. We find that GenAI exposure is highly concentrated in the city's core districts, deepening the intra-urban AI divide. Since 2023, high-exposure neighborhoods have experienced wage stagnation even as they continue to attract high-skilled workers---a ``high-skill trap.'' This wage penalty is driven by task de-skilling and intensified labor-market crowding. A difference-in-differences design centered on ChatGPT's release supports a causal interpretation. These findings challenge the prevailing theory of skill-biased technological change and provide a basis for inclusive AI governance in global technology hubs.
\end{abstract}


\newpage

% ─── Introduction ───
\section{Introduction}

Cities concentrate the workers and industries most exposed to Generative Artificial Intelligence (GenAI)~\citep{eloundou2024gpts,webb2019impact,frank2019toward}. As the first automation wave to penetrate cognitive, creative, and communicative tasks---capabilities that have long commanded an urban wage premium---GenAI raises a fundamental question: will it act as a ``great leveler''~\citep{cairncross1997death,friedman2005world} that disperses opportunity to urban peripheries~\citep{autor2015why,barrero2021why,ramani2024working}, or will it deepen core--periphery divides and widen wage gaps~\citep{florida2017new,autor2014skills,moretti2021effect}? The answer has direct implications for intra-urban digital divides, labor protections in exposed occupations, and the governance of global tech hubs~\citep{vinuesa2020role,capraro2024impact}.

\begin{figure}[htbp]
  \centering
  \includegraphics[width=0.75\textwidth]{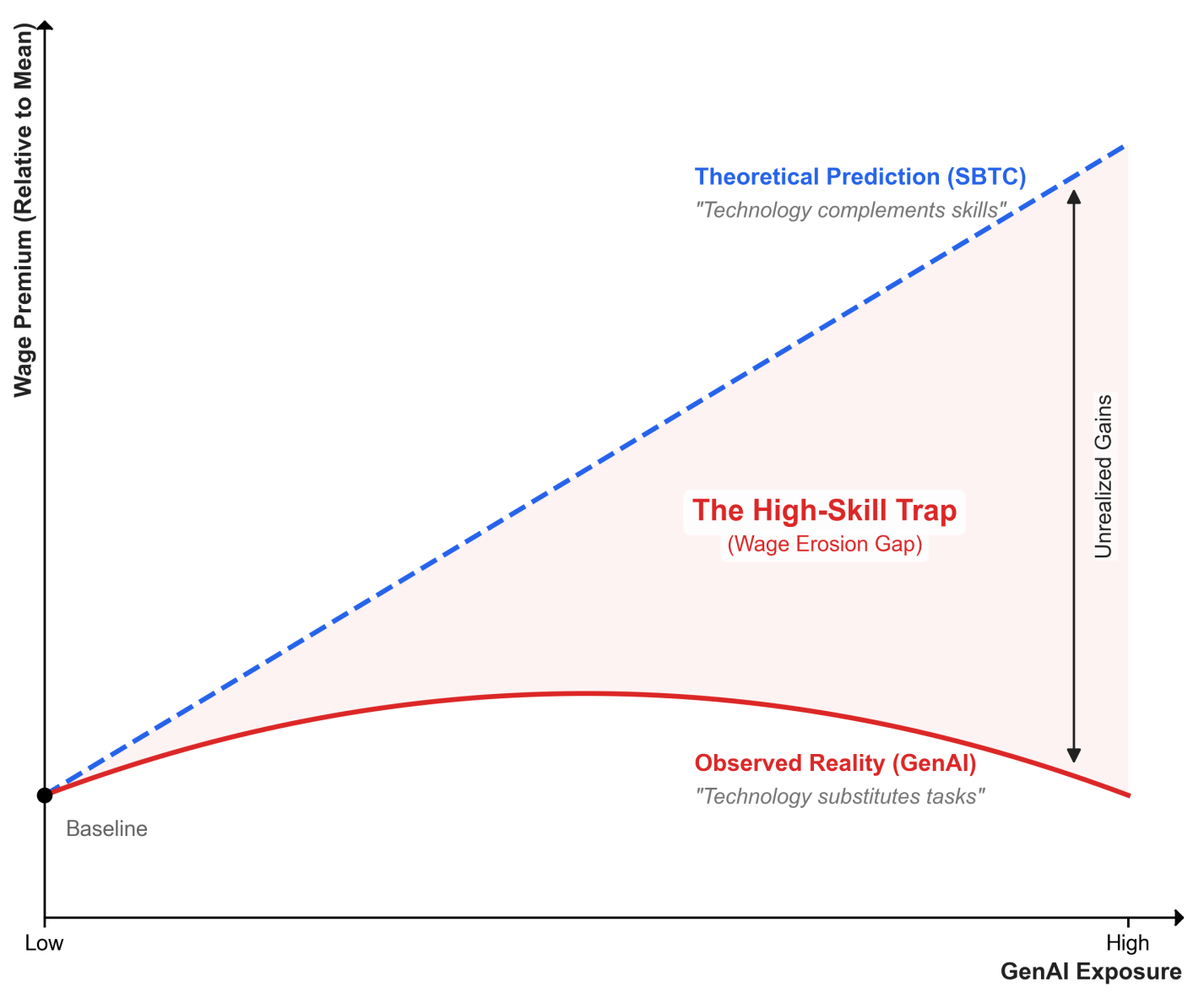}
  \caption{\textbf{The ``high-skill trap'' under GenAI exposure.} The blue dashed line shows the skill premium predicted by traditional skill-biased technological change (SBTC) theory: as exposure to technology rises, high-skilled labor and the skill premium grow linearly. The red solid line traces the pattern we observe empirically: after an initial rise, the skill premium flattens and ultimately stagnates, revealing diminishing marginal returns. The shaded region marks the ``high-skill trap''---the foregone wage gains that emerge because GenAI erodes the scarcity of high-skilled tasks, even as exposure increases.}
  \label{fig:conceptual}
\end{figure}

Previous waves of technology (Supplementary Text~S1)---from computerization to industrial robots---rewarded cognitive skills and reinforced agglomeration, creating a spatial polarization in which educated talent, high-value jobs, and rising wages clustered in city cores~\citep{moretti2012new,katz1992changes,card2002skill,barany2018job,acemoglu2011handbook,autor2003skill}. GenAI breaks this pattern. Its capabilities span the full skill spectrum, from routine coding and writing to complex data analysis, and it may compress the very skill premium that high-skilled workers have historically commanded. Three frameworks compete. The augmentation view holds that GenAI raises productivity across the skill distribution and creates new demand~\citep{brynjolfsson2025generative,noy2023experimental}. The substitution view warns of structural displacement and deepening inequality~\citep{eloundou2024gpts,acemoglu2025simple}. The task-reorganization view emphasizes that technology reshapes the content of work without wholesale replacement~\citep{frank2019toward,autor2003skill,kok2014cities}. Recent evidence points to a more counterintuitive scenario, which we call a ``high-skill trap'': a surge of skilled workers in exposed occupations coincides with stagnant wages, as GenAI erodes the scarcity value of premium cognitive tasks~\citep{eloundou2024gpts,webb2019impact,brynjolfsson2025generative,hui2024short} (Fig.~\ref{fig:conceptual}).

Three critical gaps persist. First, existing studies are predominantly national in scale, cross-sectional, and focused on high-income Western economies; they cannot reveal how GenAI reshapes intra-urban core--periphery dynamics at fine spatial resolution~\citep{frank2019toward,moretti2021effect,sun2024large}. Second, the literature lacks dynamic evidence linking exposure to synchronous changes in skill supply and wages, as well as causal evidence that exploits clear exogenous breaks---few studies leverage natural experiments such as the release of ChatGPT~\citep{eloundou2024gpts,webb2019impact}. Third, occupational exposure is typically measured using a single LLM or expert-elicitation framework~\citep{eloundou2024gpts,tolan2021measuring,acemoglu2022tasks,gmyrek2023generative}, which embeds model-specific bias. These gaps are especially consequential for global tech-hub cities, which concentrate frontier AI capital alongside heterogeneous labor markets where the stakes for distribution are highest~\citep{demombynes2025exposure,castells2011rise}.

\begin{figure}[htbp]
  \centering
  \includegraphics[width=0.9\textwidth]{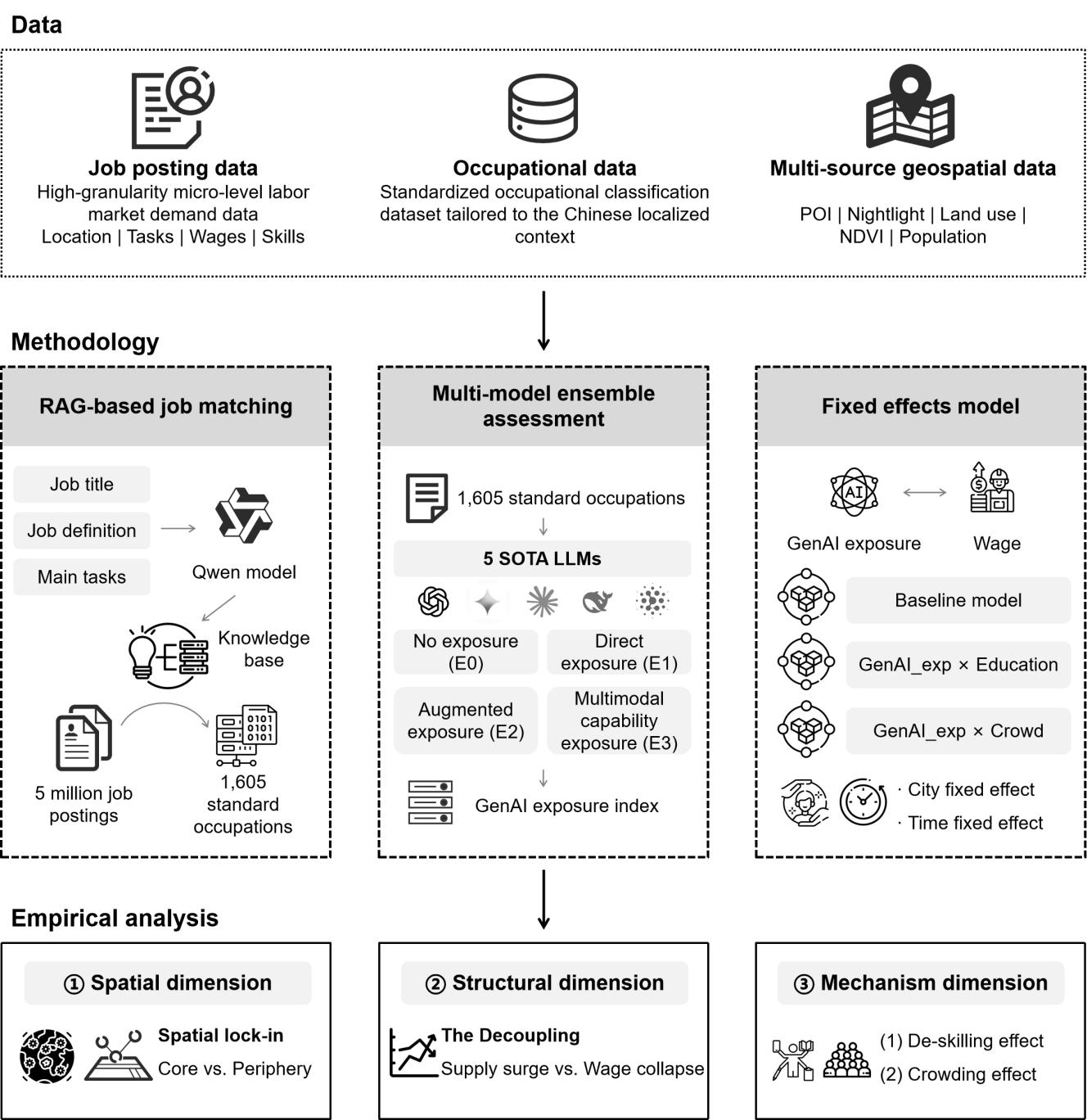}
  \caption{\textbf{Measuring GenAI exposure and its spatial, structural, and mechanistic effects.} The framework combines three steps: (1) millions of job posting are linked to standard occupational tasks using retrieval-augmented generation (RAG); (2) GenAI exposure scores are computed by aggregating assessments from five SOTA LLMs; and (3) fixed-effects econometric models are used to isolate the distinct spatial, structural, and causal mechanisms of the GenAI shock.}
  \label{fig:framework}
\end{figure}

We address these gaps using 4,995,615 online job postings from Beijing (2018--2024), matched at the neighborhood scale with multi-source geospatial data (Fig.~\ref{fig:framework}). Beijing offers a compelling setting: it hosts a globally significant AI ecosystem (Baidu, ByteDance, leading research institutions) while exhibiting a pronounced dual structure of high-tech cores and low-skill peripheries~\citep{you2026china,wu2020towards,tian2010spatial,lai2025comparative}. To measure exposure, we designed a matching pipeline that uses retrieval-augmented generation (RAG) to link job tasks to AI capabilities and then aggregated assessments from five state-of-the-art (SOTA) LLMs to reduce model-specific bias. Our analysis yields three interconnected findings. First, GenAI exposure is spatially locked into urban cores, deepening the intra-urban AI divide. Second, after ChatGPT's release, neighborhoods with high exposure experience both rising skilled-labor supply and stagnant wages---a high-skill trap. Third, this decoupling is associated with both de-skilling and labor-market crowding. To establish causality, we exploit the quasi-exogenous shock of ChatGPT's release, using pre-determined 2018 exposure as the treatment in a difference-in-differences (DID) design, with pre-trends validated through an event study.

This study makes four contributions. On measurement, we provide the first fine-grained map of GenAI exposure at the neighborhood level, opening up an intra-urban perspective largely missing from existing work. On causal identification, we pair the quasi-exogenous shock of ChatGPT's release with pre-determined 2018 exposure to construct a rigorous DID design, supported by randomization inference and a Bartik shift-share instrumental variable (IV). On mechanisms, we jointly quantify the de-skilling and crowding channels that drive the wage penalty within a unified framework. On policy, we offer an empirical foundation for inclusive AI governance in global tech-hub cities.

% ─── Figure 1: Conceptual ───

% ─── Figure 2: Framework ───

% ─── Results ───
\section{Results}

\subsection*{GenAI exposure is spatially locked into core districts}

\begin{figure}[htbp]
  \centering
  \includegraphics[width=\textwidth]{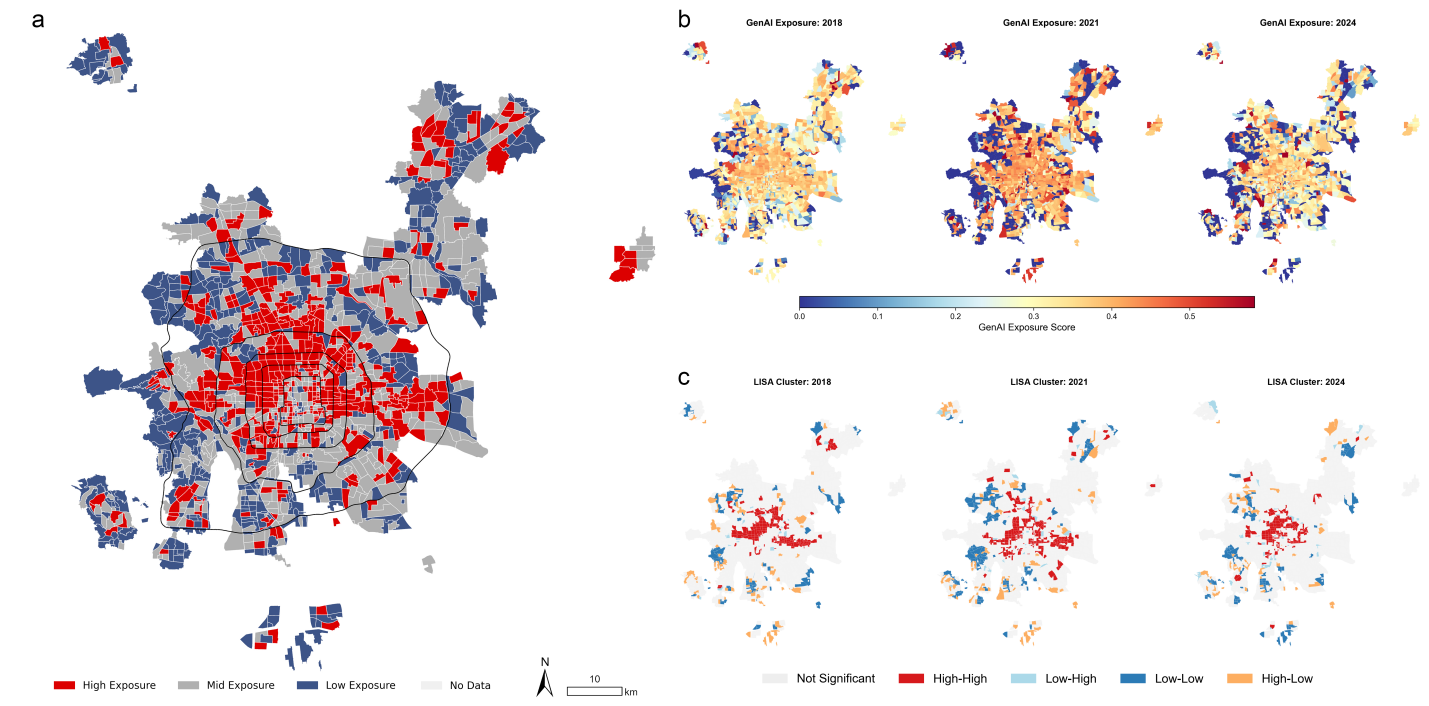}
  \caption{\textbf{GenAI exposure across Beijing: spatial patterns, temporal change, and local clustering (2018--2024).} \textbf{a,} GenAI exposure intensity averaged over 2018--2024 across Beijing's neighborhoods. \textbf{b,} Exposure distribution in 2018, 2021 and 2024, showing persistent core concentration and selective post-2023 spillover. \textbf{c,} Local spatial association (LISA) map: high--high clusters remain locked in the core with low--low clusters expanding across the periphery.}
  \label{fig:spatial}
\end{figure}

GenAI does not spread evenly across the city. Contrary to the idea that new technologies dissolve geographic barriers~\citep{cairncross1997death,friedman2005world}, we find that GenAI exposure reinforces Beijing's existing spatial hierarchy. Operating through agglomeration economies, the shock deepens the core--periphery divide and projects a new digital divide onto the urban landscape. The spatial pattern of GenAI exposure across Beijing (2018--2024) follows a monocentric, multi-ring structure with pronounced heterogeneity (Fig.~\ref{fig:spatial}a). High-exposure areas are locked into a ``golden triangle'' of knowledge-intensive services---Zhongguancun Science City (innovation), Financial Street (capital management) and the Guomao CBD (business services)---where exposure scores consistently exceed 0.3. This concentration indicates that high-cognitive, creative occupations most susceptible to AI (such as algorithm engineers and financial analysts) still depend on the face-to-face interaction and knowledge spillovers that dense urban cores provide~\citep{storper2004buzz,capello2014spatial}. In contrast, low-exposure areas are scattered across the city's ecological-conservation zones and legacy industrial districts on the periphery, where low-skill services and traditional manufacturing dominate---tasks that AI currently struggles to replicate~\citep{acemoglu2020robots}.

A temporal comparison (Fig.~\ref{fig:spatial}b) reveals two trajectories that confirm strong path dependence. First, core districts exhibit spatial lock-in and cumulative advantage: despite the pandemic and industry shocks, high-value clusters in Haidian and Chaoyang intensified rather than dispersed, with AI-related employment continuing to concentrate in talent-rich, information-dense areas~\citep{moretti2012new}. Second, the post-ChatGPT GenAI boom triggered a selective spillover. Exposure rose markedly in Beijing's municipal sub-center between 2023 and 2024, showing that emerging clusters with sufficient absorptive capacity begin to capture these spillovers. Yet this diffusion is highly selective---outer suburban counties remain low-exposure, suggesting that the spread of AI exposure is constrained by gaps in digital infrastructure and complementary assets~\citep{goldfarb2019digital} and fails to reach the periphery.

A local spatial autocorrelation analysis (Local Indicators of Spatial Association, LISA) confirms this polarization (Fig.~\ref{fig:spatial}c). High--high clusters (neighborhoods with high exposure surrounded by similarly high values) are persistently locked into the core, and their spatial connectivity strengthened between 2018 and 2024; spillovers circulate mainly within this cluster, forming a closed network of knowledge exchange. Low--low clusters (low-exposure neighborhoods surrounded by low-exposure areas) sprawl contiguously toward the southwest and northeast, creating a ``spatial poverty trap'' where workers lack both direct exposure and a supportive environment for upgrading~\citep{kraay2014poverty}. Transitional zones (high--low and low--high outliers) are largely absent, underscoring a spatial fragmentation so severe that the digital divide becomes especially difficult to bridge.

% ─── Figure 3: Spatial ───

\subsection{High-exposure neighborhoods attract talent but wages stagnate}

Spatial polarization captures the geography of the shock; structural analysis illuminates how GenAI reshapes returns to human capital. Using micro-data from 2018--2024, we examine how GenAI exposure, skill supply, and wages evolve together, testing whether the shock has structurally reversed the skill premium. The results challenge the conventional skill-biased technological change (SBTC) hypothesis~\citep{katz1992changes} and reveal a ``high-skill trap'': although high-exposure neighborhoods continue to draw in highly educated workers, they have experienced marked wage stagnation and even declines.

\begin{figure}[htbp]
  \centering
  \includegraphics[width=\textwidth]{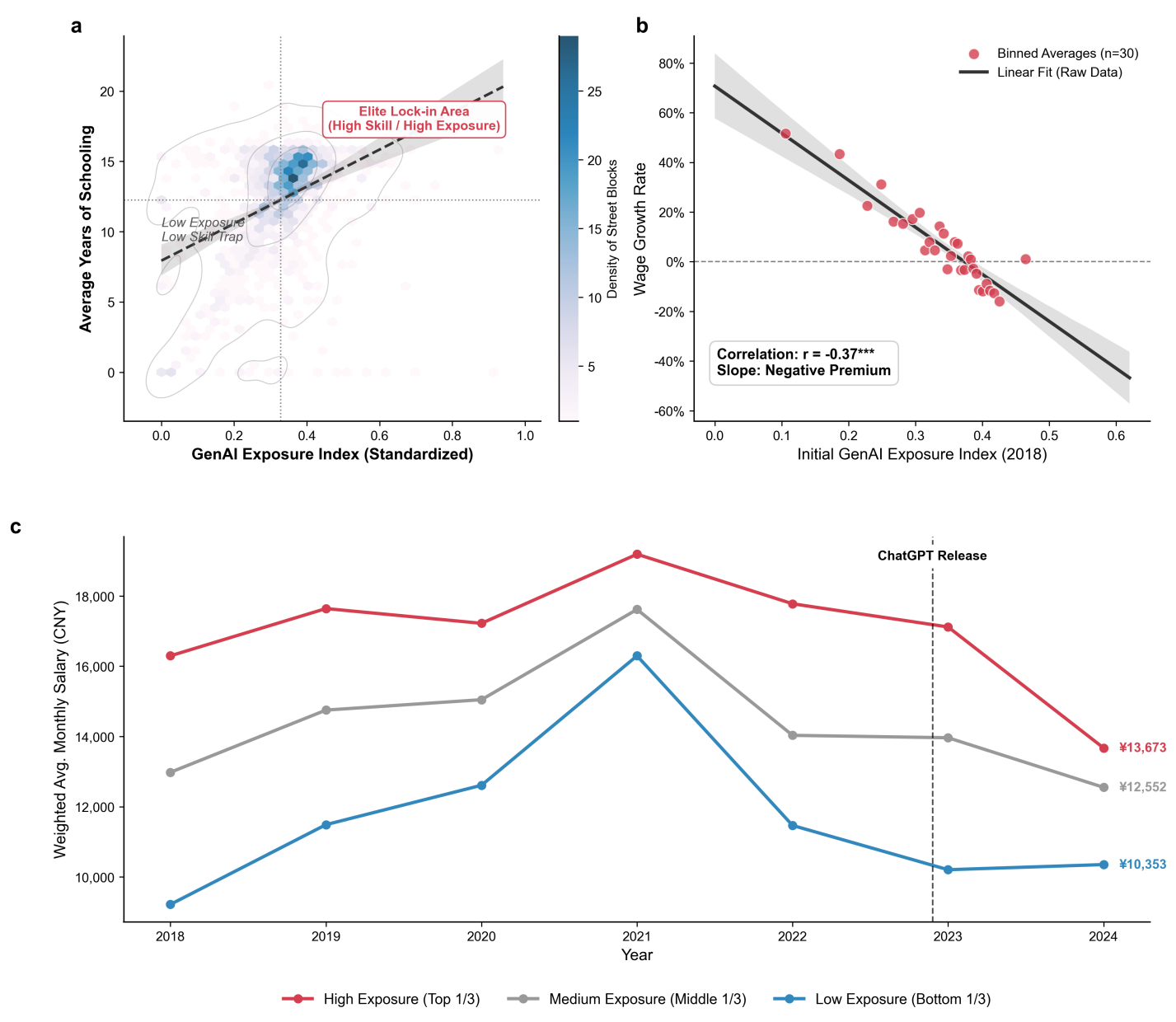}
  \caption{\textbf{GenAI exposure, education, and the emerging wage penalty.} \textbf{a,} Hexagonal heatmap of standardized GenAI exposure versus average years of schooling in 2024. The dashed line is a linear fit; the region where high education and high exposure overlap forms an ``elite lock-in'' zone. \textbf{b,} Binned scatter plot relating 2018 GenAI exposure to wage growth over 2018--2024. The negative slope (solid line) and 95\% confidence interval (shaded) indicate a wage penalty for higher initial exposure. \textbf{c,} Average monthly wages (CNY) for high, medium, and low GenAI exposure groups. The dashed line marks ChatGPT's release; after that point, the high-exposure group (red) shows a marked downturn.}
  \label{fig:wage}
\end{figure}

Fig.~\ref{fig:wage}a shows the joint distribution of average years of schooling and standardized GenAI exposure across Beijing's neighborhoods in 2024. The upward-sloping fitted line confirms a strong positive correlation: exposure rises with education. A pronounced ``elite lock-in'' zone emerges, where the workforce averages more than 12 years of schooling and GenAI exposure is at its highest---direct evidence that the GenAI shock is biased toward high-skill occupations~\citep{eloundou2024gpts}. At the other end, low-education neighborhoods fall into a low-exposure, low-skill trap: they are temporarily shielded from AI-driven displacement, but they also miss out on the potential productivity gains that AI augmentation could bring.

Yet the expected wage gains have not materialized. Fig.~\ref{fig:wage}c tracks wage trajectories for groups with high, medium, and low exposure. Wages in the high-exposure group declined sharply after peaking in 2021. Following ChatGPT's release in late 2022, they fell further to 13,673 CNY per month, narrowing the gap with the medium- and low-exposure groups. Wage in low-exposure areas, by contrast, held relatively steady---consistent with a pattern in which GenAI compresses the wage distribution primarily by pulling down high-end wages~\citep{brynjolfsson2025generative}. Fig.~\ref{fig:wage}b reinforces this picture: the relationship between initial 2018 exposure and subsequent six-year wage growth is strongly negative, indicating a clear wage penalty for high exposure. This finding revises the classical SBTC framework: in the age of GenAI, technological exposure no longer uniformly delivers a skill premium; instead, it can suppress wage growth through task substitution~\citep{acemoglu2019automation}.

% ─── Figure 4: Wage ───

\begin{figure}[htbp]
  \centering
  \includegraphics[width=\textwidth]{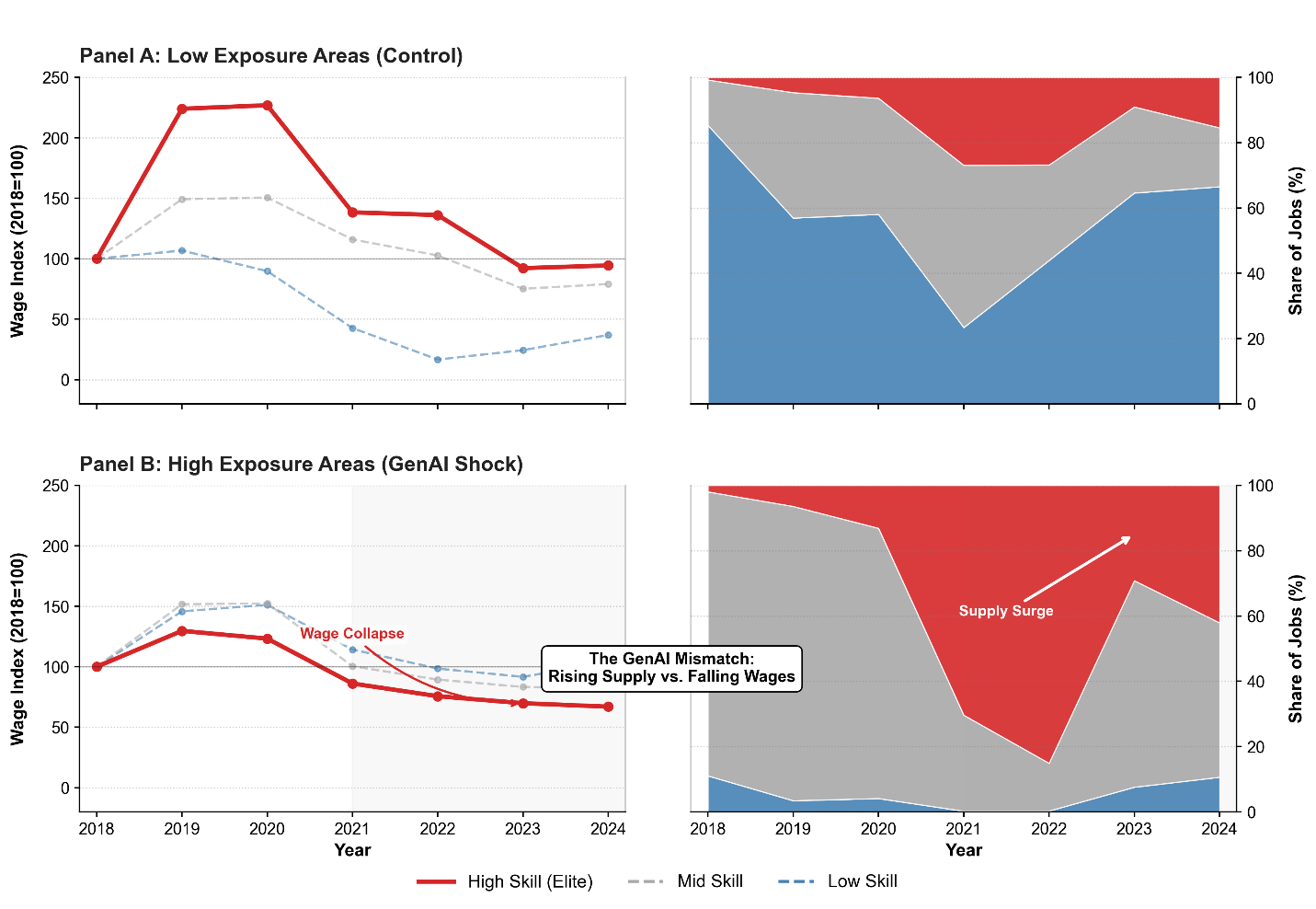}
  \caption{\textbf{The decoupling: wage stagnation versus up-skilling in high-exposure areas.} Panel~A shows low-exposure (control) areas; Panel~B shows high-exposure (treatment) areas. Lines depict the wage index and stacked areas depict the employment structure. The divergence in Panel~B---a labor-supply surge alongside a wage collapse---is the focal pattern.}
  \label{fig:decoupling}
\end{figure}

Why do neighborhoods that attract highly educated workers and concentrate high GenAI exposure still see wages stagnate? Fig.~\ref{fig:decoupling} reveals a sharp contrast between low- and high-exposure areas. In low-exposure areas, skill composition and wages move broadly together. In high-exposure areas, however, the two decouple: the share of high-skill jobs surges after 2021 as the urban core continues to draw in educated talent, yet the wage index for those same high-skill jobs declines substantially over the same period. This decoupling points to two mutually reinforcing channels. First, task substitution and de-skilling: GenAI lowers the cognitive threshold for tasks once concentrated in high-skill occupations---coding, copywriting, data analysis---making them more routine and eroding the bargaining power of skilled workers~\citep{hartley2024labor}. Second, crowding and hyper-competition: the steady inflow of educated talent into the core collides with a task space compressed by AI, generating an oversupply of workers relative to the remaining high-value tasks~\citep{bergmann1974occupational}.

% ─── Figure 5: Decoupling ───

\subsection*{Causal identification of the GenAI wage penalty}

The patterns shown so far are descriptive, but association is not causation. Between 2018 and 2024, Beijing's labor market faced overlapping shocks---antitrust tightening in the technology sector (from 2021), the COVID-19 pandemic and recovery (2020--2023), and a real-estate downturn (from 2022)---any of which could coincide spatially with GenAI exposure. To isolate the causal effect of GenAI, we use a DID design with a pre-determined treatment variable, validate the parallel-trends assumption with an event-study specification, and test robustness through randomization inference (Methods).

Two features of the research design support causal interpretation. First, the treatment is pre-determined: we use GenAI exposure in 2018, four years before ChatGPT's release (November 2022) and before our 2020--2024 analysis window, ruling out reverse causality. Second, the DID model interacts this pre-determined exposure with a post-2022 dummy indicator ($\mathrm{Post}_t$) while controlling for the three concurrent shocks---each also interacted with post-period---to separate their confounding effects from the GenAI shock. The key identifying assumption is that, in the absence of the GenAI shock, high- and low-exposure neighborhoods would have followed parallel wage trends.

\begin{figure}[htbp]
  \centering
  \includegraphics[width=0.85\textwidth]{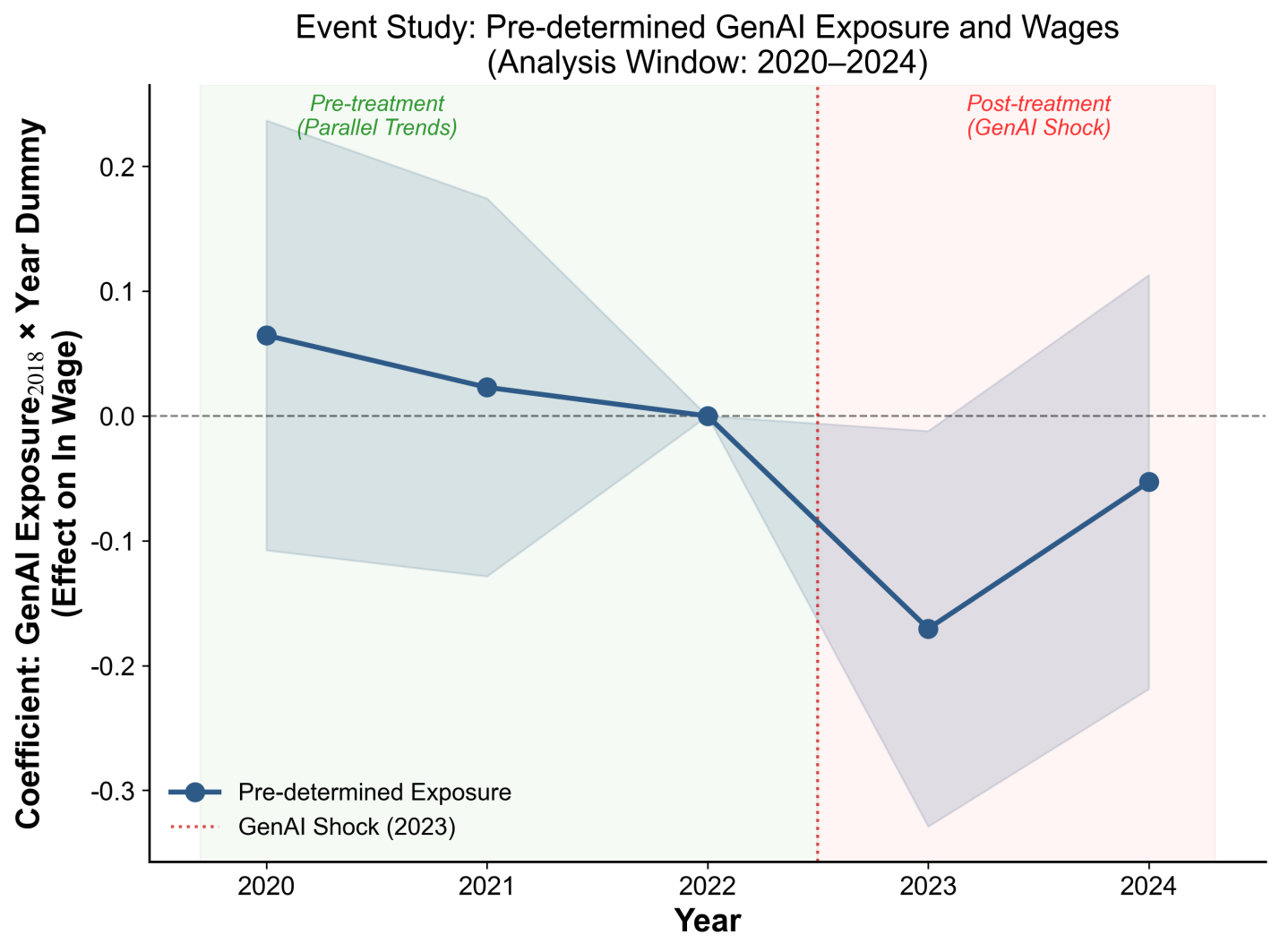}
  \caption{\textbf{Causal identification of the GenAI wage penalty.} Event-study estimates of the dynamic effect of pre-determined (2018) GenAI exposure on log wages, controlling for concurrent shocks. The reference year is 2022. Green shading marks the pre-treatment period (2020--2021) and red shading marks the post-treatment period (2023--2024). Pre-treatment coefficients are statistically indistinguishable from zero (joint Wald: $\chi^2 = 0.095$, $p = 0.916$), supporting the parallel-trends assumption. A sharp negative break emerges in 2023 ($\beta_{2023} = -0.193$, $p = 0.033$), followed by partial recovery in 2024 ($\beta_{2024} = -0.103$, $p = 0.119$). Shaded bands around point estimates represent 95\% confidence intervals based on neighborhood-clustered standard errors.}
  \label{fig:eventstudy}
\end{figure}

Fig.~\ref{fig:eventstudy} reports the event-study estimates with the concurrent-shock controls (2022 is the reference year; Equation~\ref{eq:eventstudy}). Pre-treatment coefficients ($\beta_{2020} = 0.065$, $p = 0.462$; $\beta_{2021} = 0.023$, $p = 0.767$) are statistically insignificant. A joint Wald test fails to reject the parallel trends assumption ($\chi^2 = 0.095$, $p = 0.916$). A sharp structural break appears post-treatment. The $\beta_{2023}$ coefficient drops to $-0.193$ ($p = 0.033$), implying that a one-standard-deviation increase in pre-determined exposure is associated with an approximately 17.5\% decline in wages during the first year of the shock. The 2024 coefficient is $-0.103$ ($p = 0.119$), still negative but with weakened significance, consistent with partial labor-market adjustment through occupational transitions, skill restructuring, or the gradual emergence of new AI-complementary roles.

Table~\ref{tab:did} reports the core DID estimates. In the baseline (Column~1), the interaction coefficient is $-0.140$ ($p < 0.05$), equivalent to a wage decline of about 13.1\% post-shock. When the three concurrent-shock controls are added (Column~2), the coefficient does not shrink; it increases slightly to $-0.151$ ($p < 0.05$). This stability is informative: if the concurrent shocks were driving the result, adding them as controls would have reduced the estimate; instead, the estimate remains robust, suggesting that any omitted-variable bias from those shocks is more likely to mask the GenAI effect than to create it spuriously~\citep{oster2019unobservable}. None of the interactions with the concurrent shocks are statistically significant, ruling out the possibility that those shocks independently account for the findings.

We further test the robustness of the DID results with randomization inference. Shuffling the pre-determined exposure measure randomly across neighborhoods 500 times produces a distribution of placebo estimates tightly centered on zero. The true estimate ($-0.151$) lies in the extreme left tail of this distribution (permutation $p = 0.004$; only 2 of 500 simulations yielded a more extreme coefficient). This confirms that the wage penalty is neither noise nor an artefact of data structure.

As complementary evidence, we construct a Bartik shift-share IV that predicts local GenAI exposure using national AI-related job growth weighted by each neighborhood's 2018 industrial composition (Equations~\ref{eq:bartik1}--\ref{eq:reducedform}). The reduced-form estimate from this instrument implies that a one-standard-deviation increase in the Bartik predictor reduces wage growth by 0.436 log points ($p < 0.001$), which aligns closely with the DID estimate. The first-stage F-statistic of 53.5 far exceeds the weak-instrument threshold~\citep{stock2002testing}. However, because the Bartik event-study fails the parallel-trends test (joint Wald $p = 0.004$), the structural parameter should be interpreted with caution. Full diagnostics results appear in Supplementary Text~S6.

% ─── Table 1: DID ───
\begin{table}[htbp]
  \centering
  \caption{\textbf{Difference-in-Differences Estimates of the GenAI Wage Penalty.} The dependent variable is the natural logarithm of average monthly wages. $\mathrm{GenAI}_{2018}$ denotes standardized pre-determined GenAI exposure measured in 2018. Post~=~1 for 2023--2024. The analysis window spans 2020--2024 with 2022 as the base year. Confounder controls are industry-composition-based exposure measures for technology regulation, COVID recovery and real-estate adjustment, each interacted with Post. Control variables include ln(population), nightlight intensity, NDVI, POI density and land-use ratio. Standard errors clustered at the neighborhood level are in parentheses. **~$p < 0.05$; *~$p < 0.1$.}
  \label{tab:did}
  \small
  \begin{tabular}{lcc}
    \toprule
    & (1) Baseline DID & (2) + Confounder controls \\
    \midrule
    Dependent variable: $\ln(\mathrm{Wage})$ & & \\
    $\mathrm{GenAI}_{2018} \times \mathrm{Post}$ & $-0.140^{**}$ & $-0.151^{**}$ \\
     & $(0.061)$ & $(0.068)$ \\[4pt]
    $\mathrm{TechReg} \times \mathrm{Post}$ & & $0.035$ \\
     & & $(0.061)$ \\[4pt]
    $\mathrm{COVID} \times \mathrm{Post}$ & & $0.092$ \\
     & & $(0.087)$ \\[4pt]
    $\mathrm{RealEstate} \times \mathrm{Post}$ & & $-0.039$ \\
     & & $(0.109)$ \\[4pt]
    Controls & Yes & Yes \\
    Entity FE & Yes & Yes \\
    Time FE & Yes & Yes \\
    Observations & 6,895 & 6,895 \\
    Neighborhoods & 1,383 & 1,383 \\
    Pre-trend joint test ($p$) & 0.758 & 0.916 \\
    Permutation $p$-value & 0.004 & 0.004 \\
    \bottomrule
  \end{tabular}
\end{table}

% ─── Figure 6: Event study ───

\subsection*{De-skilling and crowding erode the skill premium}

Classical SBTC theory predicts that new technologies complement high-skilled workers, widening the skill premium~\citep{goldin2018race}. Recent work on GenAI points to the opposite mechanism---de-skilling, or skill-levelling. The idea is that GenAI lowers the cognitive barriers to tasks once reserved for high-skilled workers (such as coding, copywriting, and data analysis), allowing lower-skilled workers to substitute for high-skilled labor and compressing wages at the top~\citep{acemoglu2019automation,hartley2024labor,dellacqua2023navigating}. We test this directly by examining how education moderates the wage effect of GenAI exposure.

Column~2 of Table~\ref{tab:mechanism} shows that, although the main effect of education on wages remains significantly positive (1.150), the interaction between GenAI and education is negative and highly significant ($-1.286$). This means that the extra wage gain associated with GenAI exposure shrinks as education rises. Supplementary Fig.~S1a reveals two regimes. For low-education workers, GenAI exposure has a positive marginal effect---consistent with an assistive role in which AI helps workers clear skill-specific task barriers~\citep{brynjolfsson2025generative}. As education increases, however, this effect declines and eventually turns negative for the most highly educated workers, including those with master's degrees and above. For this elite cohort, GenAI's capabilities overlap substantially with their own human capital (programming, translation, writing), and the technology acts more as a substitute than a complement, eroding their bargaining power. These results provide strong support for the de-skilling hypothesis~\citep{webb2019impact,dellacqua2023navigating}.

Skill substitution alone does not explain the full wage penalty. A second, complementary channel operates through labor-market crowding. According to the crowding hypothesis~\citep{bergmann1974occupational}, when the supply of workers in AI-exposed fields grows rapidly and the demand for their tasks does not expand enough, productivity gains can paradoxically push wages down---a pattern often called ``involution.'' We capture this dynamic using a measure of job-market heat, which reflects the intensity of competition for positions in each occupation. Column~3 of Table~\ref{tab:mechanism} reports a highly significant negative interaction between GenAI exposure and job-market heat ($-1.619$, $p < 0.01$), indicating that greater competitive intensity strongly suppresses the GenAI skill premium. Supplementary Fig.~S1b traces three distinct phases. In low-competition niches, the marginal effect of GenAI exposure is positive, generating a ``blue ocean'' of excess returns through scarcity. As adoption diffuses and competition approaches the mean, the premium shrinks and crosses zero, marking an equilibrium phase. In the most competitive segments---``red ocean'' phase---the marginal effect turns sharply negative, confirming a crowding-driven wage penalty.

Together, these findings reveal a technology-induced saturation mechanism: widespread GenAI adoption lowers barriers and homogenizes the skill signals that once differentiated high-skilled workers, intensifying competition and compressing wages. To move from panel associations to causal evidence, we estimate a triple-DID specification that interacts pre-determined exposure, the post-2022 period, and either education or job-market heat. The triple interaction coefficients are $-0.135$ ($p = 0.178$) for the de-skilling channel and $-0.063$ ($p = 0.436$) for the crowding channel. Both signs are consistent with the panel results. Given the short post-treatment window, neither reaches conventional significance, but their directional consistency offers preliminary causal support (Supplementary Text~S7).

% ─── Table 2: Mechanism ───
\begin{table}[htbp]
  \centering
  \caption{\textbf{Mechanisms underlying the GenAI wage effect.} The dependent variable is the natural logarithm of monthly wages. Column~1 reports the baseline two-way fixed-effects specification. Column~2 introduces the interaction between GenAI exposure and education to test the de-skilling channel. Column~3 introduces the interaction between GenAI exposure and job-market heat to test the crowding channel. Standard errors (in parentheses) are clustered at the neighborhood level. Significance levels: ***~$p < 0.01$, **~$p < 0.05$, *~$p < 0.1$. Variance inflation factors are below 5 for all variables, ruling out serious multicollinearity. While the DID analysis is restricted to the 2020--2024 window to allow parallel-trends verification, these regressions use the full 2018--2024 panel ($N = 9{,}653$ neighborhood--year observations) to maximize statistical power for estimating conditional correlations and their heterogeneity.}
  \label{tab:mechanism}
  \small
  \begin{tabular}{lccc}
    \toprule
    & (1) Baseline & (2) De-skilling & (3) Crowding \\
    \midrule
    Dep.\ Var.: $\ln(\mathrm{Wage})$ & & & \\[2pt]
    GenAI Exposure (Std.) & $0.835^{***}$ & $0.800^{***}$ & $-0.233^{*}$ \\
     & $(0.068)$ & $(0.087)$ & $(0.138)$ \\[4pt]
    Education (Std.) & $1.699^{***}$ & $1.150^{***}$ & $1.369^{***}$ \\
     & $(0.120)$ & $(0.099)$ & $(0.074)$ \\[4pt]
    Job Market Heat (Std.) & $1.736^{***}$ & $1.194^{***}$ & $1.633^{***}$ \\
     & $(0.168)$ & $(0.176)$ & $(0.260)$ \\[4pt]
    \multicolumn{4}{l}{\textit{Interactions}} \\[2pt]
    GenAI $\times$ Education & & $-1.286^{***}$ & \\
     & & $(0.053)$ & \\[4pt]
    GenAI $\times$ Job Market Heat & & & $-1.619^{***}$ \\
     & & & $(0.088)$ \\[4pt]
    \multicolumn{4}{l}{\textit{Controls}} \\[2pt]
    Population (ln) & $-101.250^{***}$ & $-71.956^{**}$ & $-71.868^{***}$ \\
     & $(31.637)$ & $(29.076)$ & $(27.459)$ \\[4pt]
    Nightlight & $0.147$ & $0.164$ & $0.113$ \\
     & $(0.143)$ & $(0.116)$ & $(0.115)$ \\[4pt]
    NDVI (Std.) & $0.219$ & $0.126$ & $0.024$ \\
     & $(0.208)$ & $(0.111)$ & $(0.147)$ \\[4pt]
    POI Density (Std.) & $0.034$ & $-0.006$ & $-0.087$ \\
     & $(0.100)$ & $(0.049)$ & $(0.061)$ \\[4pt]
    Land Use Ratio (Std.) & $0.433$ & $0.174$ & $0.208$ \\
     & $(0.385)$ & $(0.284)$ & $(0.249)$ \\[4pt]
    \multicolumn{4}{l}{\textit{Model Statistics}} \\[2pt]
    Entity FE & Yes & Yes & Yes \\
    Time FE & Yes & Yes & Yes \\
    Observations & 9,653 & 9,653 & 9,653 \\
    $R^2$ & 0.613 & 0.747 & 0.723 \\
    $F$-statistic & 1636.2 & 2706.8 & 2397.1 \\
    \bottomrule
  \end{tabular}
\end{table}

% ─── Discussion ───
\section{Discussion}

This study uses fine-grained data from Beijing (2018--2024) to reveal a triple polarization---spatial, structural and wage-related---that GenAI is producing in a global megacity. Our first finding challenges the classic ``death of distance'' hypothesis, which predicts that technologies that slash the cost of moving information will disperse high-skilled work geographically~\citep{cairncross1997death}. In reality, GenAI-exposed jobs remain tightly clustered in the ``golden triangle'' of Zhongguancun, Financial Street, and the CBD, and this clustering did not weaken after ChatGPT's release---if anything, it intensified. We attribute this paradox to the changing nature of knowledge in the GenAI era. While codified information and routine analysis can now flow almost frictionlessly through large models, high-value creative work still relies on tacit knowledge---the kind of know-how that is exchanged through face-to-face interaction, informal social networks and localized innovation ecosystems~\citep{storper2004buzz,glaeser2012triumph,polanyi2009knowledge}. GenAI, rather than dissolving the need for proximity, acts as a complementary asset that magnifies the returns to agglomeration, leaving peripheral neighborhoods that lack this cognitive infrastructure at deepening disadvantage~\citep{moretti2012new,goldfarb2019digital}.

More fundamentally, our results document the breakdown of SBTC theory in the context of GenAI. The long-standing view that new technologies inherently favor high-skilled workers and therefore widen the skill premium~\citep{acemoglu2011handbook,acemoglu2022tasks,goldin2018race,autor2024new} does not hold up against the evidence. Instead, we identify a new de-skilling mechanism: GenAI exerts a unique ``top-squeeze'' effect on the most highly educated workers, consistent with the levelling effect noted by others~\citep{brynjolfsson2025generative}. It simultaneously lowers the barriers for routine coding, writing, and analytical tasks, allowing lower-skilled workers to move into previously high-skilled roles. On the other hand, it directly encroaches on the core cognitive tasks traditionally embedded in high-skilled occupations, rapidly diluting their scarcity~\citep{acemoglu2018race}. This explains why the supply of educated workers and their wages have decoupled in high-exposure neighborhoods. Accumulating academic credentials no longer guarantees rising returns; the logic of returns to education itself is being restructured~\citep{webb2019impact,brynjolfsson2025generative,deming2021growing}. Wage stagnation, however, is not solely the product of substitution. It is amplified by labor-market crowding. In popular, AI-exposed fields, widespread adoption intensifies competition: as GenAI raises baseline output expectations, workers must invest more effort simply to hold their position~\citep{dellacqua2023navigating}. When collective upskilling is not met by expanding demand for the newly accessible tasks, it fails to translate into productivity gains and instead weakens workers' bargaining power~\citep{acemoglu2019automation,lazear2004internal}.

Our findings resonate with recent global reports. A 2026 study from Anthropic concludes that actual AI adoption lags behind its theoretical potential but notes a 14\% drop in hiring for the 22--25-year-old cohort~\citep{massenkoff2026labor}. Our spatial micro-data reveal that this decline is concentrated in the high-exposure core neighborhoods, pointing to strong spatial heterogeneity in AI's youth-employment effects. Similarly, the IMF~(2026)~\citep{jaumotte2026bridging} observes that AI skills command a wage premium yet simultaneously reduces entry-level hiring by 3.6\%. Our identification of the ``high-skill trap'' provides an intra-urban mechanism for that macro pattern: AI does not eliminate jobs wholesale; rather, it lowers the barriers to cognitive tasks, triggering excessive competition in spatially clustered cores.

The policy implications are clear. In responding to GenAI, policymakers should look beyond computing infrastructure or generic skills training. We highlight three directions. First, to loosen the technological monopoly of urban cores, AI demonstration zones could be established in municipal sub-centers and suburban new towns, accompanied by lower access cost for digital infrastructure. Second, higher education should pivot away from cultivating ``codifiable'' standard cognitive skills---which AI replicates readily---toward complex collaboration, critical reasoning, and cross-cultural communication, capabilities that are harder to automate~\citep{deming2017growing}. Third, labor protection policies such as re-employment transition funds and skill-transformation subsidies targeted at highly exposed occupations can prevent the wage erosion of the ``high-skill trap'' from cascading into talent outflows.

This study has limitations. The data come from Beijing alone; caution is needed when generalizing to other global cities or smaller Chinese cities. Online recruitment data capture hiring flows rather than the stock of employed workers and may over-represent volatility~\citep{hershbein2018do}. Finally, because GenAI has been widely deployed for only a short period, longer-run productivity effects---including the possible emergence of entirely new occupational categories---may not yet be visible. Tracking these evolutionary dynamics over longer horizons is a priority for future work.

% ─── Methods ───
\section{Methods}

\subsection*{Study area}

We selected Beijing as the study area for two reasons. First, it sits at the technological frontier. Beijing is a global hub for AI research and industry, concentrating China's leading AI talent, core research institutions (such as Tsinghua University and the Beijing Academy of Artificial Intelligence), and the headquarters of firms like Baidu and ByteDance. This high density of AI activity makes Beijing an early adopter of GenAI, and the resulting labor-market disruptions offers an early signal of the dynamics that may soon appear in other leading AI cities---San Francisco, London, and Singapore among them~\citep{you2026china,wu2020towards,lai2025comparative}. Second, Beijing exhibits a pronounced spatial duality. The city has a textbook core--periphery structure, combining world-class knowledge hubs such as Zhongguancun Science City with peripheral areas that retain traditional industrial or agricultural features~\citep{tian2010spatial}. This sharp spatial contrast provides an ideal setting to test whether GenAI functions as an ``equalizer'' that disperses opportunity to the periphery or as a force that deepens intra-urban divides~\citep{capello2014spatial,graham2002bridging}.

\subsection*{Core recruitment indicators and control variables}

We use online job postings from the RESSET Enterprise Big Data Platform, which aggregates vacancies from major Chinese recruitment platforms (51job, Zhaopin, BOSS Zhipin and Liepin). After cleaning---retaining only the first unique posting per company per month and dropping records with missing key fields---we spatially matched the remaining postings to Beijing's neighborhoods (regulatory planning units). The final dataset contains 4,995,615 valid recruitment postings from 1 January 2018 to 31 December 2024. From these micro-data we constructed three core neighborhood-level measures. (1)~Hiring activity. We compute the share of citywide postings located in each neighborhood (Job Share) to capture its relative position in the market, and a normalized log-transformed count (Job Index) to measure cross-neighborhood variation in employment intensity. (2)~Average monthly wage. All compensation components---annual salary, 13th-month pay, and bonuses---are converted to a monthly equivalent, winsorized at the 0.1\% tails (excluding values below 1,000 or above 280,000 RMB), and averaged within each neighborhood. (3)~Average educational requirement. We map each posting's education requirement to years of schooling (unrestricted~=~0, junior middle school or below~=~9, high school/vocational/technical~=~12, associate degree~=~15, bachelor's~=~16, master's~=~19, doctorate~=~23) and average these values within neighborhoods to capture the local skill talent threshold and degree of industrial sophistication.

To control for confounding influences, we include five categories of neighborhood-level characteristics: living convenience, development intensity, environmental quality, economic activity, and population distribution. All controls are compiled as an annual panel for 2018--2024 at the neighborhood scale. Full indicator definitions, calculation methods, and data sources appear in Supplementary Table~S2.

\subsection*{RAG-based job matching method}

Online job titles and descriptions are often unstructured and highly variable, so traditional keyword or fuzzy matching yields low accuracy. We therefore developed a standardized occupational matching framework based on RAG, achieving precise mapping between large-scale job postings and China's 2022 Occupational Classification Dictionary. First, we built a national-standard occupational knowledge base by extracting core fields---names, definitions and primary tasks---for all detailed occupations in the Dictionary and vectorizing this textual information using the Qwen3-Embedding-4B model, producing a high-dimensional semantic vector knowledge base of 1,605 standard occupations. Second, in the matching phase, the same embedding model converted job titles and responsibilities from online recruitment data into query vectors, which were then matched by semantic-similarity retrieval against the knowledge base, assigning each non-standardized posting to the closest standard occupational sub-category. To validate the method, we randomly sampled 500 matches and had them manually checked by experts. Classification accuracy exceeded 93\%, substantially outperforming traditional dictionary and fuzzy-matching approaches. Compared with end-to-end LLM-based generative classification, the RAG pipeline achieves similarly high accuracy at a fraction of the computational cost.

\subsection*{Multi-model ensemble assessment of GenAI exposure}

Using the same 1,605 standard occupations defined above (excluding party/government organs and military personnel), we constructed an assessment matrix using five SOTA LLMs---ChatGPT-4o, Gemini-2.5-pro, Claude-3.5-Sonnet, GLM-4 and Deepseek-R1---to measure the GenAI shock and to mitigate the training-data biases and hallucinations inherent in any single model. Following the assessment criteria of Eloundou et al.\ (2024)~\citep{eloundou2024gpts} and Chen et al.\ (2025)~\citep{chen2025large}, we designed standardized prompt engineering requiring each model to independently classify each occupation as exhibiting No Exposure (E0), Direct Exposure (E1), Application-Augmented Exposure (E2) or Multimodal Capability Exposure (E3). To minimize random error, we conducted five independent assessment rounds (25 inferences total) and used the mean as the final exposure score (Supplementary Text~S4). The occupational-level GenAI exposure index was then obtained through aggregation. Assessments were completed in June 2025. Pairwise correlations among model assessments range from 0.66 to 0.86, indicating consistent ensemble behavior. A manual expert review of 100 randomly sampled occupations yielded 89\% consistency with model assessments, confirming the robustness and reliability of the ensemble approach.

\subsection*{Pre-determined DID}

To establish a causal link between GenAI exposure and wage changes, we adopt a pre-determined DID design. The core specification is:
\begin{equation}
  \ln W_{it} = \alpha + \beta \, \mathrm{GenAI}_{i,2018} \times \mathrm{Post}_t + \boldsymbol{\gamma}' \mathbf{Z}_{i,2018} \times \mathrm{Post}_t + \boldsymbol{\delta}' \mathbf{X}_{it} + \mu_i + \lambda_t + \varepsilon_{it}
  \label{eq:did}
\end{equation}
where $\mathrm{GenAI}_{i,2018}$ is the standardized pre-determined GenAI exposure measured in 2018; $\mathrm{Post}_t$ is the post-treatment dummy; $\mathbf{Z}_{i,2018}$ denotes the three concurrent-shock exposures based on the 2018 industrial structure; $\mathbf{X}_{it}$ includes ln(population), nightlight intensity, NDVI, POI density and land-use ratio; and $\mu_i$ and $\lambda_t$ are neighborhood and year fixed effects, respectively. The analysis window spans 2020--2024, comprising three pre-treatment years (2020, 2021, 2022) and two post-treatment years (2023, 2024), totalling 6,895 neighborhood--year observations across 1,383 neighborhoods. Standard errors are clustered at the neighborhood level. The treatment variable's strict pre-determination is central: GenAI exposure is measured in 2018, four years before ChatGPT's release (November 2022) and prior to the analysis window (from 2020), ruling out reverse causality and anticipatory responses. The core parameter $\beta$ estimates the differential change in wages following the ChatGPT shock per standard-deviation increase in pre-determined GenAI exposure, conditional on concurrent-shock controls.

\subsection*{Confounder exposure construction}

The three concurrent-shock exposures are constructed within a unified framework. For each shock (technology regulation, COVID-19 recovery, real-estate adjustment), we identify the set of directly affected industries and compute each neighborhood's 2018 employment share in that set:
\begin{equation}
  \mathrm{Exposure}_{i}^{\mathrm{shock}} = \frac{\sum_{j \in \mathcal{J}^{\mathrm{shock}}} n_{ij,2018}}{\sum_{j} n_{ij,2018}}
  \label{eq:confounder}
\end{equation}
where $n_{ij,2018}$ is the number of recruitment postings for neighborhood $i$ in industry $j$ in 2018. Technology-regulation exposure uses the employment share of the Information Technology and Software industry (Industry Code~I); COVID-19-recovery exposure uses the combined share of Accommodation and Catering~(H) and Transportation~(G); real-estate-adjustment exposure uses the combined share of Real Estate~(K) and Construction~(E). All three exposures are standardized and interacted with $\mathrm{Post}_t$ before inclusion.

\subsection*{Event-study specification}

To verify parallel trends and characterize the dynamic effect, we estimate the event-study model:
\begin{equation}
  \ln W_{it} = \alpha + \sum_{\tau \neq 2022} \beta_\tau \, \mathrm{GenAI}_{i,2018} \times \mathbb{1}[t=\tau] + \boldsymbol{\gamma}' \mathbf{Z}_{i,2018} \times \mathrm{Post}_t + \boldsymbol{\delta}' \mathbf{X}_{it} + \mu_i + \lambda_t + \varepsilon_{it}
  \label{eq:eventstudy}
\end{equation}
with 2022 as the base period. The identifying assumption requires that all pre-treatment coefficients satisfy $\beta_\tau = 0$, meaning that prior to the GenAI shock, neighborhoods with different exposure levels followed a common wage trend. We assess this assumption with a joint Wald test.

\subsection*{Randomization inference}

To evaluate the robustness of statistical inference without relying on large-sample asymptotics or specific error-structure assumptions, we implement randomization inference grounded in Fisher's exact test. The procedure is: (1)~randomly permute pre-determined GenAI exposure across the 1,383 neighborhoods, breaking the actual mapping between treatment and outcome; (2)~re-estimate the DID coefficient in Equation~\ref{eq:did} with concurrent-shock controls using the permuted exposures; (3)~repeat 500 times to generate a placebo distribution; and (4)~compute the two-sided permutation $p$-value. The procedure is exact in finite samples and robust to arbitrary spatial correlation and heteroskedasticity.

\subsection*{Bartik shift-share IV}

As a complementary identification strategy, we construct a Bartik shift-share IV. For each of the 18 primary industries, we compute a leave-one-out industry-average GenAI exposure that excludes the focal neighborhood's contribution:
\begin{equation}
  \bar{E}_{-i,j} = \frac{\sum_{k \neq i} n_{kj} \cdot E_k}{\sum_{k \neq i} n_{kj}}
  \label{eq:bartik1}
\end{equation}
where $n_{kj}$ is the number of postings in neighborhood $k$ for industry $j$ and $E_k$ is the GenAI exposure of neighborhood $k$. Interacting neighborhood $i$'s 2018 industry-employment share with the industry-level exposure and summing over industries yields:
\begin{equation}
  \mathrm{Bartik}_i = \sum_{j} s_{ij,2018} \cdot \bar{E}_{-i,j}
  \label{eq:bartik2}
\end{equation}
This IV is driven solely by historical industrial structure and national-level industry-specific exposure, excluding endogenous adjustments by the focal neighborhood.

To maximize first-stage power, IV estimation employs a broader window than the event study. Using 2018--2019 averages as the pre-period and 2023--2024 averages as the post-period, we compute long differences ($\Delta$) for each variable and estimate cross-sectional two-stage least squares (2SLS):

First stage:
\begin{equation}
  \Delta \mathrm{GenAI}_i = \pi_0 + \pi_1 \, \mathrm{Bartik}_i + \boldsymbol{\pi}_2' \Delta \mathbf{W}_i + u_i
  \label{eq:firststage}
\end{equation}

Second stage:
\begin{equation}
  \Delta \ln W_i = \beta_0 + \beta_1 \, \widehat{\Delta \mathrm{GenAI}}_i + \boldsymbol{\beta}_2' \Delta \mathbf{W}_i + v_i
  \label{eq:secondstage}
\end{equation}

Reduced form:
\begin{equation}
  \Delta \ln W_i = \rho_0 + \rho_1 \, \mathrm{Bartik}_i + \boldsymbol{\rho}_2' \Delta \mathbf{W}_i + \eta_i
  \label{eq:reducedform}
\end{equation}
where $\mathbf{W}_i$ includes population change, nightlight change and the three concurrent-shock exposures. Instrument strength is evaluated via the first-stage $F$-statistic, benchmarked against the critical value of 10 of Stock and Yogo (2002)~\citep{stock2002testing}.

\subsection*{Triple-DID mechanism tests}

To test the de-skilling and crowding mechanisms within a causal framework, we extend the DID model into a triple-DID design. Taking the de-skilling channel as an example:
\begin{equation}
  \ln W_{it} = \alpha + \beta_1 D_i \times \mathrm{Post}_t + \beta_2 M_i \times \mathrm{Post}_t + \beta_3 D_i \times M_i \times \mathrm{Post}_t + \boldsymbol{\theta}' \mathbf{C}_{it} + \mu_i + \lambda_t + \varepsilon_{it}
  \label{eq:tripledid}
\end{equation}
where $M_i$ is the 2018 moderator (education level or recruitment heat), and $\mathbf{C}_{it}$ collects concurrent-shock controls, baseline controls and neighborhood and year fixed effects. The key parameter $\beta_3$ measures heterogeneity in the GenAI wage penalty---specifically, whether it is exacerbated by higher education (or greater market crowding). All moderators use 2018 pre-determined values to preserve exogeneity.

\subsection*{Fixed-effects model}

To examine the impact of GenAI exposure on labor-market compensation and employment structure, we estimate a two-way fixed-effects model that controls for time-invariant individual heterogeneity and individual-invariant time trends, mitigating omitted-variable bias. To uncover the de-skilling and crowding mechanisms, we add interaction terms between GenAI exposure and education level, and between GenAI exposure and job-market heat, respectively:
\begin{equation}
  \ln W_{it} = \alpha + \beta_1 G_{it} + \beta_2 E_{it} + \beta_3 G_{it} \times E_{it} + \boldsymbol{\delta}' \mathbf{X}_{it} + \mu_i + \lambda_t + \varepsilon_{it}
  \label{eq:fe_deskilling}
\end{equation}
\begin{equation}
  \ln W_{it} = \alpha + \beta_1 G_{it} + \beta_2 H_{it} + \beta_3 G_{it} \times H_{it} + \boldsymbol{\delta}' \mathbf{X}_{it} + \mu_i + \lambda_t + \varepsilon_{it}
  \label{eq:fe_crowding}
\end{equation}
where $G_{it}$ is standardized GenAI exposure; $E_{it}$ is the standardized education level; $H_{it}$ is standardized recruitment heat (market crowding); $\beta_3$ tests the de-skilling effect (Eq.~\ref{eq:fe_deskilling}) and crowding effect (Eq.~\ref{eq:fe_crowding}); $\mathbf{X}_{it}$ is a vector of controls covering living convenience, development intensity, environmental quality, economic activity and population distribution; and $\mu_i$ and $\lambda_t$ are entity and time fixed effects.

To assess the robustness of the interaction-term coefficients, we conducted a placebo test based on random permutation: the GenAI exposure variable was randomly shuffled across neighborhood--year observations, the shuffled variable was used to reconstruct the interaction terms, and the placebo coefficients were re-estimated. Repeating this procedure 500 times yielded an empirical distribution of placebo coefficients. The fact that the true interaction coefficient lies far in the tail of this distribution indicates that the observed de-skilling and crowding effects are not statistical noise (Supplementary Fig.~S6 and Supplementary Text~S8).

% ─── Data & Code Availability ───
\section{Data availability}

The datasets generated and analyzed during the current study are not publicly available because they contain location-identifiable spatial information that raises privacy concerns, but they are available from the corresponding author on reasonable request.

\section{Code availability}

Code underlying this study is not publicly available but may be made available to qualified researchers on reasonable request from the corresponding author.

% ─── Author contributions ───
\section*{Author contributions}

\noindent X.H.: conceptualization, methodology, software, validation, formal analysis, investigation, data curation, writing---original draft, visualization, project administration.

\noindent H.Z.: methodology, investigation, software, writing---review and editing.

\noindent M.M.: methodology, visualization, writing---review and editing.

\noindent E.W.L.: methodology, data curation, writing---review and editing.

\noindent K.E.: methodology, software, writing---review and editing.

\noindent A.H.: writing---review and editing.

\noindent J.L.: writing---review and editing.

\noindent L.C.: data curation, writing---review and editing.

\noindent Y.L.: conceptualization, methodology, resources, writing---review and editing, supervision, funding acquisition.

% ─── Competing interests ───
\section*{Competing interests}

The authors declare no competing interests.

% ─── References ───
\bibliographystyle{apalike}

\begin{thebibliography}{58}

\bibitem[Acemoglu(2025)]{acemoglu2025simple}
Acemoglu, D. The simple macroeconomics of AI. \textit{Economic Policy} 40, 13--58 (2025).

\bibitem[Acemoglu and Autor(2011)]{acemoglu2011handbook}
Acemoglu, D. \& Autor, D. Skills, tasks and technologies: Implications for employment and earnings. In \textit{Handbook of Labor Economics} Vol.~4, 1043--1171 (Elsevier, 2011).

\bibitem[Acemoglu and Restrepo(2018)]{acemoglu2018race}
Acemoglu, D. \& Restrepo, P. The race between man and machine: Implications of technology for growth, factor shares, and employment. \textit{American Economic Review} 108, 1488--1542 (2018).

\bibitem[Acemoglu and Restrepo(2019)]{acemoglu2019automation}
Acemoglu, D. \& Restrepo, P. Automation and new tasks: How technology displaces and reinstates labor. \textit{Journal of Economic Perspectives} 33, 3--30 (2019).

\bibitem[Acemoglu and Restrepo(2020)]{acemoglu2020robots}
Acemoglu, D. \& Restrepo, P. Robots and jobs: Evidence from US labor markets. \textit{Journal of Political Economy} 128, 2188--2244 (2020).

\bibitem[Acemoglu and Restrepo(2022)]{acemoglu2022tasks}
Acemoglu, D. \& Restrepo, P. Tasks, automation, and the rise in US wage inequality. \textit{Econometrica} 90, 1973--2016 (2022).

\bibitem[Autor(2014)]{autor2014skills}
Autor, D. H. Skills, education, and the rise of earnings inequality among the ``other 99 percent.'' \textit{Science} 344, 843--851 (2014).

\bibitem[Autor(2015)]{autor2015why}
Autor, D. H. Why are there still so many jobs? The history and future of workplace automation. \textit{Journal of Economic Perspectives} 29, 3--30 (2015).

\bibitem[Autor et al.(2003)]{autor2003skill}
Autor, D. H., Levy, F. \& Murnane, R. J. The skill content of recent technological change: An empirical exploration. \textit{The Quarterly Journal of Economics} 118, 1279--1333 (2003).

\bibitem[Autor et al.(2024)]{autor2024new}
Autor, D., Chin, C., Salomons, A. \& Seegmiller, B. New frontiers: The origins and content of new work, 1940--2018. \textit{The Quarterly Journal of Economics} 139, 1399--1465 (2024).

\bibitem[B\'ar\'any and Siegel(2018)]{barany2018job}
B\'ar\'any, Z. L. \& Siegel, C. Job polarization and structural change. \textit{American Economic Journal: Macroeconomics} 10, 57--89 (2018).

\bibitem[Barrero et al.(2021)]{barrero2021why}
Barrero, J. M., Bloom, N. \& Davis, S. J. Why working from home will stick. NBER Working Paper No.~28731 (2021).

\bibitem[Bergmann(1974)]{bergmann1974occupational}
Bergmann, B. R. Occupational segregation, wages and profits when employers discriminate by race or sex. \textit{Eastern Economic Journal} 1, 103--110 (1974).

\bibitem[Brynjolfsson et al.(2025)]{brynjolfsson2025generative}
Brynjolfsson, E., Li, D. \& Raymond, L. Generative AI at work. \textit{The Quarterly Journal of Economics} 140, 889--942 (2025).

\bibitem[Cairncross(1997)]{cairncross1997death}
Cairncross, F. \textit{The Death of Distance: How the Communications Revolution Will Change Our Lives} (Harvard Business School Press, 1997).

\bibitem[Capello and Lenzi(2014)]{capello2014spatial}
Capello, R. \& Lenzi, C. Spatial heterogeneity in knowledge, innovation, and economic growth nexus. \textit{Journal of Regional Science} 54, 186--214 (2014).

\bibitem[Capraro et al.(2024)]{capraro2024impact}
Capraro, V. et al. The impact of generative artificial intelligence on socioeconomic inequalities and policy making. \textit{PNAS Nexus} 3, pgae191 (2024).

\bibitem[Card and DiNardo(2002)]{card2002skill}
Card, D. \& DiNardo, J. E. Skill-biased technological change and rising wage inequality: Some problems and puzzles. \textit{Journal of Labor Economics} 20, 733--783 (2002).

\bibitem[Castells(2011)]{castells2011rise}
Castells, M. \textit{The Rise of the Network Society} 2nd edn (John Wiley \& Sons, 2011).

\bibitem[Chen et al.(2025)]{chen2025large}
Chen, Q., Ge, J., Xie, H., Xu, X. \& Yang, Y. Large language models at work in China's labor market. \textit{China Economic Review} 92, 102413 (2025).

\bibitem[Dell'Acqua et al.(2023)]{dellacqua2023navigating}
Dell'Acqua, F. et al. Navigating the jagged technological frontier: Field experimental evidence of the effects of AI on knowledge worker productivity and quality. \textit{Harvard Business School Technology \& Operations Mgt. Unit Working Paper} (2023).

\bibitem[Deming(2017)]{deming2017growing}
Deming, D. J. The growing importance of social skills in the labor market. \textit{The Quarterly Journal of Economics} 132, 1593--1640 (2017).

\bibitem[Deming(2021)]{deming2021growing}
Deming, D. J. The growing importance of decision-making on the job. NBER Working Paper No.~29193 (2021).

\bibitem[Demombynes et al.(2025)]{demombynes2025exposure}
Demombynes, G., Langbein, J. \& Weber, M. The exposure of workers to Artificial Intelligence in low-and middle-income countries. World Bank Policy Research Working Paper (2025).

\bibitem[Eloundou et al.(2024)]{eloundou2024gpts}
Eloundou, T., Manning, S., Mishkin, P. \& Rock, D. GPTs are GPTs: Labor market impact potential of LLMs. \textit{Science} 384, 1306--1308 (2024).

\bibitem[Florida(2017)]{florida2017new}
Florida, R. \textit{The New Urban Crisis: How Our Cities Are Increasing Inequality, Deepening Segregation, and Failing the Middle Class---and What We Can Do About It} (Basic Books, 2017).

\bibitem[Frank et al.(2019)]{frank2019toward}
Frank, M. R. et al. Toward understanding the impact of artificial intelligence on labor. \textit{Proceedings of the National Academy of Sciences} 116, 6531--6539 (2019).

\bibitem[Friedman(2005)]{friedman2005world}
Friedman, T. L. \textit{The World Is Flat: A Brief History of the Twenty-First Century} (Macmillan, 2005).

\bibitem[Glaeser(2012)]{glaeser2012triumph}
Glaeser, E. \textit{Triumph of the City: How Our Greatest Invention Makes Us Richer, Smarter, Greener, Healthier, and Happier} (Penguin, 2012).

\bibitem[Goldfarb and Tucker(2019)]{goldfarb2019digital}
Goldfarb, A. \& Tucker, C. Digital economics. \textit{Journal of Economic Literature} 57, 3--43 (2019).

\bibitem[Goldin and Katz(2018)]{goldin2018race}
Goldin, C. \& Katz, L. F. The race between education and technology. In \textit{Inequality in the 21st Century} 49--54 (Routledge, 2018).

\bibitem[Gmyrek et al.(2023)]{gmyrek2023generative}
Gmyrek, P., Berg, J. \& Bescond, D. Generative AI and jobs: A global analysis of potential effects on job quantity and quality. ILO Working Paper (2023).

\bibitem[Graham(2002)]{graham2002bridging}
Graham, S. Bridging urban digital divides? Urban polarisation and information and communications technologies (ICTs). \textit{Urban Studies} 39, 33--56 (2002).

\bibitem[Hartley et al.(2024)]{hartley2024labor}
Hartley, J., Jolevski, F., Melo, V. \& Moore, B. The labor market effects of generative artificial intelligence. Available at SSRN (2024).

\bibitem[Hershbein and Kahn(2018)]{hershbein2018do}
Hershbein, B. \& Kahn, L. B. Do recessions accelerate routine-biased technological change? Evidence from vacancy postings. \textit{American Economic Review} 108, 1737--1772 (2018).

\bibitem[Hui et al.(2024)]{hui2024short}
Hui, X., Reshef, O. \& Zhou, L. The short-term effects of generative artificial intelligence on employment: Evidence from an online labor market. \textit{Organization Science} 35, 1977--1989 (2024).

\bibitem[Jaumotte et al.(2026)]{jaumotte2026bridging}
Jaumotte, M. F. et al. Bridging skill gaps for the future: New jobs creation in the AI age. IMF Working Paper (2026).

\bibitem[Katz and Murphy(1992)]{katz1992changes}
Katz, L. F. \& Murphy, K. M. Changes in relative wages, 1963--1987: Supply and demand factors. \textit{The Quarterly Journal of Economics} 107, 35--78 (1992).

\bibitem[Kok and Weel(2014)]{kok2014cities}
Kok, S. \& Weel, B. t. Cities, tasks, and skills. \textit{Journal of Regional Science} 54, 856--892 (2014).

\bibitem[Kraay and McKenzie(2014)]{kraay2014poverty}
Kraay, A. \& McKenzie, D. Do poverty traps exist? Assessing the evidence. \textit{Journal of Economic Perspectives} 28, 127--148 (2014).

\bibitem[Lai and Zhao(2025)]{lai2025comparative}
Lai, Y. \& Zhao, H. Comparative analysis of smart city scientific research trends in the USA and China. \textit{Nature Cities} 2, 875--883 (2025).

\bibitem[Lazear and Oyer(2004)]{lazear2004internal}
Lazear, E. P. \& Oyer, P. Internal and external labor markets: A personnel economics approach. \textit{Labour Economics} 11, 527--554 (2004).

\bibitem[Massenkoff and McCrory(2026)]{massenkoff2026labor}
Massenkoff, M. \& McCrory, P. Labor market impacts of AI: A new measure and early evidence. \textit{Anthropic Research} 5 (2026).

\bibitem[Moretti(2012)]{moretti2012new}
Moretti, E. \textit{The New Geography of Jobs} (Houghton Mifflin Harcourt, 2012).

\bibitem[Moretti(2021)]{moretti2021effect}
Moretti, E. The effect of high-tech clusters on the productivity of top inventors. \textit{American Economic Review} 111, 3328--3375 (2021).

\bibitem[Noy and Zhang(2023)]{noy2023experimental}
Noy, S. \& Zhang, W. Experimental evidence on the productivity effects of generative artificial intelligence. \textit{Science} 381, 187--192 (2023).

\bibitem[Oster(2019)]{oster2019unobservable}
Oster, E. Unobservable selection and coefficient stability: Theory and evidence. \textit{Journal of Business \& Economic Statistics} 37, 187--204 (2019).

\bibitem[Polanyi(2009)]{polanyi2009knowledge}
Polanyi, M. \textit{The Tacit Dimension} In \textit{Knowledge in Organisations} 135--146 (Routledge, 2009).

\bibitem[Ramani et al.(2024)]{ramani2024working}
Ramani, A., Alcedo, J. \& Bloom, N. How working from home reshapes cities. \textit{Proceedings of the National Academy of Sciences} 121, e2408930121 (2024).

\bibitem[Stock and Yogo(2002)]{stock2002testing}
Stock, J. H. \& Yogo, M. Testing for weak instruments in linear IV regression. NBER Working Paper No.~T0284 (2002).

\bibitem[Storper and Venables(2004)]{storper2004buzz}
Storper, M. \& Venables, A. J. Buzz: Face-to-face contact and the urban economy. \textit{Journal of Economic Geography} 4, 351--370 (2004).

\bibitem[Sun et al.(2024)]{sun2024large}
Sun, Y., Zhu, H., Wang, L., Zhang, L. \& Xiong, H. Large-scale online job search behaviors reveal labor market shifts amid COVID-19. \textit{Nature Cities} 1, 150--163 (2024).

\bibitem[Tian et al.(2010)]{tian2010spatial}
Tian, G., Wu, J. \& Yang, Z. Spatial pattern of urban functions in the Beijing metropolitan region. \textit{Habitat International} 34, 249--255 (2010).

\bibitem[Tolan et al.(2021)]{tolan2021measuring}
Tolan, S. et al. Measuring the occupational impact of AI: Tasks, cognitive abilities and AI benchmarks. \textit{Journal of Artificial Intelligence Research} 71, 191--236 (2021).

\bibitem[Vinuesa et al.(2020)]{vinuesa2020role}
Vinuesa, R. et al. The role of artificial intelligence in achieving the Sustainable Development Goals. \textit{Nature Communications} 11, 233 (2020).

\bibitem[Webb(2019)]{webb2019impact}
Webb, M. The impact of artificial intelligence on the labor market. Available at SSRN 3482150 (2019).

\bibitem[Wu et al.(2020)]{wu2020towards}
Wu, F. et al. Towards a new generation of artificial intelligence in China. \textit{Nature Machine Intelligence} 2, 312--316 (2020).

\bibitem[You(2026)]{you2026china}
You, X. China intensifies push to become world leader in tech and AI. \textit{Nature} 651, 859--860 (2026).

\end{thebibliography}

\end{document}